\documentclass{article}
\usepackage{graphicx} 

\usepackage{amssymb}
\usepackage{amsmath}
\usepackage{amsthm}
\usepackage{authblk}
\usepackage{colortbl}
\usepackage{multirow}
\usepackage{comment}
\usepackage{bm}
\usepackage{algorithm}
\usepackage{algpseudocode}

\usepackage[margin=13truemm]{geometry}

\usepackage{url}

\title{Clifford+V synthesis for multi-qubit unitary gates}
\author[1]{Soichiro Yamazaki}
\author[2,3]{Seiseki Akibue}
\affil[1]{\footnotesize The University of Tokyo, 7-3-1 Hongo, Bunkyo, Tokyo 113-8654, Japan.}
\affil[2]{\footnotesize Communication Science Laboratories, NTT, Inc., 3-1 Morinosato Wakamiya, Atsugi, Kanagawa 243-0198, Japan.}
\affil[3]{\footnotesize NTT Research Center for Theoretical Quantum Information, NTT, Inc., 3--1, Morinosato Wakamiya, Atsugi, Kanagawa 243-0198, Japan.}

\date{}

\begin{document}
\maketitle

\begin{abstract}
We developed a general framework for synthesizing target gates by using a finite set of basic gates, which is a crucial step in quantum compilation. When approximating a gate in SU($n$), a naive brute-force search requires a computational complexity of $O(1/\varepsilon^{(n^2 - 1)})$ to achieve an approximation with error $\varepsilon$. In contrast, by using our method, the complexity can be reduced to $O(-n^2 \log\varepsilon/\varepsilon^{((n^2 - 1)/2)})$. This method requires almost no assumptions and can be applied to a variety of gate sets, including Clifford+$T$ and Clifford+$V$. Further, we introduce a suboptimal but short run-time algorithm for synthesizing multi-qubit controlled gates. This approach highlights the role of subgroup structures in reducing synthesis complexity and opens a new direction of study on the compilation of multi-qubit gates. The framework is broadly applicable to different universal gate sets, and our analysis suggests that it can serve as a foundation for resource-efficient quantum compilation in near-term architectures.
\end{abstract}

\section{Introduction}

Fault-tolerant quantum computation (FTQC) enables reliable quantum processing by encoding logical qubits into physical ones using quantum error correction (QEC). In this framework, every gate operation is carried out as a logical gate to protect against decoherence. Many QEC codes, particularly stabilizer codes, facilitate efficient implementation of Clifford gates due to their transversal nature. However, as established by the Gottesman-Knill theorem \cite{G98}, quantum circuits composed solely of Clifford gates can be efficiently simulated on classical computers, and thus do not offer a quantum advantage.

To achieve universal quantum computation, it is necessary to introduce at least one non-Clifford gate—such as the T gate, the V gate, or the controlled-S gate—into the gate set. Although methods for implementing these gates at the logical level using magic states have been developed \cite{BK05,DS13}, they incur significantly higher resource costs compared with Clifford gates. This disparity has driven extensive research into optimizing the number of non-Clifford gates required to implement a given unitary operation. The process of decomposing a target unitary operation into a sequence of gates from a specific elementary gate set, such as Clifford + T or Clifford + V, is called unitary synthesis or quantum compilation.

Unitary synthesis often aims to approximate a desired unitary operation within a desired approximation error while minimizing the number of costly non-Clifford gates. This is because many quantum algorithms rely on unitary operations having continuous parameters, while only discrete sets of logical gates are available. It is important to emphasize that this approximation issue does not arise from a particular selection of QEC codes; rather, it is an unavoidable consequence when logical gates comprise a finite number of non-transversal gates in conjunction with transversal gates \cite{EK09}.

While synthesis algorithms for the Clifford + T gate set have been the most extensively studied, increasing attention has been given to Clifford + V gate synthesis. These elementary gate sets share a deep connection with number theory, allowing for common algorithm design. In particular, various synthesis algorithms have been proposed for Clifford + V circuits, including a Pauli rotation and a general single-qubit unitary gate \cite{PhysRevA.88.012313}.

Despite these advances, synthesizing multi-qubit unitaries remains a formidable challenge. Even for large approximation errors (e.g., $\epsilon \simeq 10^{-2}$), identifying the optimal gate sequence becomes computationally intractable. The best-known algorithms \cite{PhysRevA.109.052619} for this task have almost the same time complexity as that of a brute-force search, with the search space size growing exponentially in the number of V gates. This underscores the need for more efficient synthesis algorithms in practical multi-qubit applications.

\begin{table}
\centering
\begin{tabular}{|c||c||c|c|c|} \hline
    Synthesizing method & Gate & $R_z$ & 1-qubit unitary & $N$-qubit unitary \\
    \hline\hline
    Matsumoto \& Amano (2008) & $T$ & \begin{tabular}{c}
        $3\log_2(1/\epsilon)$ \\ $O(1/\epsilon^3)$
    \end{tabular} & \begin{tabular}{c}
        $3\log_2(1/\epsilon)$ \\ $O(1/\epsilon^3)$
    \end{tabular} & \textit{$N^24^N$ 1-qubit unitaries} \\
    \hline
    Ross \& Selinger (2016) & $T$ & \begin{tabular}{c}
        $3\log_2(1/\epsilon)$ \\ $poly(\log(1/\epsilon))$
    \end{tabular} & \begin{tabular}{c}
        $9\log_2(1/\epsilon)$ \\ $poly(\log(1/\epsilon))$
    \end{tabular} & \textit{$N^24^N$ 1-qubit unitaries} \\
    \hline
    Gheorghiu et al. (2023) & $T$ & \begin{tabular}{c}
        $3\log_2(1/\epsilon)$ \\ $O(1/\epsilon^3)$
    \end{tabular} & \begin{tabular}{c}
        $3\log_2(1/\epsilon)$ \\ $O(1/\epsilon^3)$
    \end{tabular} & \begin{tabular}{c}
        $\geq(4^N-1)\log_{4^N}(1/\epsilon)$ \\ $O(4^{NT_\epsilon+2N})$
    \end{tabular}\\
    \hline
    This paper & $T$ & \begin{tabular}{c}
        $3\log_2(1/\epsilon)$ \\ $O(\log(1/\epsilon)/\epsilon^{1.5})$
    \end{tabular} & \begin{tabular}{c}
        $3\log_2(1/\epsilon)$ \\ $O(\log(1/\epsilon)/\epsilon^{1.5})$
    \end{tabular} & \begin{tabular}{c}
        $\geq(4^N-1)\log_{2\cdot 4^N-3}(1/\varepsilon)$ \\ $O((NT_\epsilon)2^{NT_\epsilon})$
    \end{tabular}\\
    \hline
    Bocharov et al. (2013) & $V$ & \begin{tabular}{c}
        $3\log_5(1/\epsilon)$ \\ $poly(\log(1/\epsilon))$
    \end{tabular} & \begin{tabular}{c}
        $3\log_5(1/\epsilon)$ \\ $O(1/\epsilon)$
    \end{tabular} & \textit{$N^24^N$ 1-qubit unitaries}\\
    \hline
    Mukhopadhyay (2024) & $V$ & \begin{tabular}{c}
        $3\log_5(1/\epsilon)$ \\ $O(1/\epsilon^3)$
    \end{tabular} & \begin{tabular}{c}
        $3\log_5(1/\epsilon)$ \\ $O(1/\epsilon^3)$
    \end{tabular} & \begin{tabular}{c}
        $\geq(4^N-1)\log_{2\cdot 4^N-3}(1/\varepsilon)$ \\ $O((2\cdot 4^N-3)^{V_\epsilon})$
    \end{tabular}\\
    \hline
    This paper & $V$ & \begin{tabular}{c}
        $3\log_5(1/\epsilon)$ \\ $O(\log(1/\epsilon)/\epsilon^{1.5})$
    \end{tabular} & \begin{tabular}{c}
        $3\log_5(1/\epsilon)$ \\ $O(\log(1/\epsilon)/\epsilon^{1.5})$
    \end{tabular} & \begin{tabular}{c}
        $\geq(4^N-1)\log_{2\cdot 4^N-3}(1/\varepsilon)$ \\ $O(V_\epsilon(2\cdot 4^N-3)^{V_\epsilon/2})$
    \end{tabular}\\
    \hline
\end{tabular}
\caption{Comparison of Clifford+$T$ and $V$ synthesizing methods. The upper half of the figure shows the method of synthesis using Clifford+$T$, while the lower half shows the method using Clifford+$V$. Within each box, the top part indicates the number of gates (excluding constant overhead) after synthesis, while the bottom represents the time complexity of the synthesis. In the boxes for the N-qubit unitary, the time complexity is written using $T_\epsilon$ and $V_\epsilon$ to denote the $T$ count and $V$ count. }
\end{table}

\subsection{Our contribution}
We propose two new synthesis algorithms that improve achievable approximating error for multi-qubit unitary operations. First, we introduce a method called meet-in-the-middle exhaustive search, which enables the search of the optimal gate sequences up to double the length of brute-force approaches, thereby achieving an approximation error scaling quadratically better per fixed execution time. Furthermore, by combining with the technique of probabilistic synthesis \cite{SA}, our method achieves an approximation error scaling biquadratically better than the deterministic synthesis based on brute-force approaches. This algorithm combines the meet-in-the-middle technique with nearest neighbor search (NNS) (e.g. kd-tree \cite{10.1145/361002.361007}) or approximate nearest neighbor search (ANNS) (e.g. Hierarchical Navigable Small World (HNSW) \cite{malkov2018efficient}, Navigating Spread-out Graph (NSG) \cite{FuNSG17}) data structure to efficiently search large gate spaces for synthesizing general multi-qubit unitary operations.
 To the best of our knowledge, the application of meet-in-the-middle techniques has been limited to exact synthesis scenarios \cite{PhysRevA.109.052619}.
Furthermore, it can be easily generalized to other universal gate sets. This can be done by simply changing the basis to be multiplied. Since there is no restriction on the gate sets, this algorithm can be safely said to be a generalized one.
In addition, because the precomputation of partial products and the nearest-neighbor queries are independent, the method naturally lends itself to parallel computation, which would further improve scalability on modern hardware.

Second, we present a suboptimal synthesis algorithm specialized for multi-qubit-controlled-unitary operations, which are central to numerous quantum algorithms, including the quantum Fourier transform (QFT) \cite{C02}, quantum singular value transformation (QSVT) \cite{LC19,GSLW19}, and Hamiltonian simulation \cite{AN12}. While this algorithm does not necessarily provide the optimal decomposition, in the 2-qubit case, it reduces the V-count by 30~\% compared with the conventional method that reduces the synthesis of multi-qubit unitary operations into that of single-qubit unitary operations \cite{NielsenChuang}. Moreover, in the 2-qubit case, it can achieve a 60~\% smaller approximation log-error within the same runtime relative to a naive application of meet-in-the-middle exhaustive search. 
The innovative aspect of this synthesis algorithm lies in our identification of a relatively large set of unitaries with V-count 1 such that it uniformly covers a subgroup of multi-qubit unitary operations.
This also enables us to generalize the target to the following type of gates for the $N$-qubit case.
\begin{equation}
    U_1\oplus U_2\oplus \cdots\oplus U_{2^{N-1}}\qquad (U_1,U_2,\cdots,U_{2^{N-1}}\in\mathrm{SU(2)})
\end{equation}
In this generalized case, the V-count is reduced by 40~\% compared with that of composition of single-qubit gates, and the log-error is decreased by 70~\% relative to the direct application of the meet-in-the-middle method.

\section{Preliminaries}
\subsection{V-basis}
In this paper, the V-basis for an $n$-qubit system is composed of the following gates:
\begin{equation}
    \left\{\frac{\mathbb{I}\pm2iP_{\bm{x}}}{\sqrt{5}} \; \middle| \;\bm{x}\in\{0,1,2,3\}^n\setminus\{(0,0,\cdots,0)\}, P_{\bm{x}}=P_{x_0}\otimes P_{x_1}\otimes\cdots\otimes P_{x_{n-1}}\right\},
\end{equation}
while $P_0=\mathbb{I}$, $P_1=X$, $P_2 = Y$, and $P_3 = Z$.
For a 1-qubit system, the following three gates and their conjugates form a V-basis:
\begin{equation}
    V_x = \frac{\mathbb{I}+2iX}{\sqrt{5}},~V_y = \frac{\mathbb{I}+2iY}{\sqrt{5}},~V_z = \frac{\mathbb{I}+2iZ}{\sqrt{5}}.
\end{equation}
Every V-basis is generated only by using a 1-qubit V-basis and Clifford gates, and every gate generated from a 1-qubit V-basis and Clifford gates is included in a V-basis.
Because the V-basis is closed under conjugation by Clifford gates, any Clifford+V sequence can be expressed, without altering the number of V-basis elements, as a Clifford gate followed by a product of V-basis elements.

\subsection{V-count}
The V-count is the number of $V$ gates in a line of Clifford + V gates.

Given a target unitary $U\in\mathrm{SU}(2^n)$, let $\mathcal{V}_\epsilon(U)$ denote the minimum V-count for synthesizing a unitary $U^\prime$ that satisfies $|\mathcal{U}(U)-\mathcal{U}(U^\prime)|_\diamond < \epsilon$, where $\mathcal{U}(U)$ is a completely positive and trace preserving (CPTP) map defined as $\mathcal{U}(U)(\rho):=U\rho U^\dagger$ and $|\mathcal{A}-\mathcal{B}|_\diamond$ is the diamond norm between CPTP maps $\mathcal{A}$ and $\mathcal{B}$. 
For the case of a single qubit, a lower bound on the average V-count is $3\log_5(1/\epsilon)+\mathrm{constant}$ because the target unitaries are distributed in a 3-sphere, while the number of exactly implementable gates scales as $5^n$ for V-count $n$.
$U\in\mathrm{SU(2)}$ can be decomposed as
\begin{equation} \label{eq:su2_4d}
    U=a\mathbb{I}+biX+ciY+diZ
\end{equation}
where $a,b,c,d\in\mathbb{R}$ and $a^2+b^2+c^2+d^2=1$.
Moreover, it has been experimentally verified that this V-count is achievable in most cases \cite{PhysRevA.88.012313}.
From here on, the constant after the logarithm will be ignored.
In the case of 2-qubit, a lower bound on the average V-count is $15\log_{29}(1/\epsilon)$.
Similarly, a lower bound on the average V-count in the $n$-qubit case is $(4^n-1)\log_{2\cdot 4^n-3}(1/\epsilon)$. Although these lower bounds are known, it has not been determined whether they are attainable values when $n\geq2$.

The minimum V-counts can be achieved by performing a direct search, which takes $O(1/\epsilon^{4^n-1})$ time naively.
(A fast $O(1/\epsilon)$ algorithm is known for 1-qubit case \cite{PhysRevA.88.012313}.)
This naive method takes a lot of time, and it would only be a dream that it could achieve $\epsilon < 10^{-5}$.
Here, what is needed is a fast algorithm that enables an $n\geq2$-qubit gate Clifford + V decomposition even at the expense of the optimal V-count.

Since calculating the diamond norm is an expensive procedure, we utilized the phase-invariant norm for SU($d$) instead:
\begin{equation}
    |U-V| = \sqrt{1-\frac{|\mathrm{tr}(UV^\dagger)|}{d}}.
\end{equation}

\section{Methods}
\subsection{Conditionally controlled gate synthesis} \label{ssec:cgs}
We dub
\begin{equation}
    \begin{pmatrix}
        A\\&B
    \end{pmatrix}
    \in \mathrm{SU(4)}
\end{equation}
as $C(A,B)$.
For the $n>2$-qubit case, we have
\begin{equation}
    C(A_1, A_2,\cdots,A_{2^{n-1}}) = \begin{pmatrix}
        A_1\\
        &A_2\\
        &&\ddots\\
        &&&A_{2^{n-1}}
    \end{pmatrix}\in\mathrm{SU(2^n)}.
\end{equation}

Let us consider $C(A,B)$ where $A$ and $B$ are not explicitly targeted but the following $U$ is targeted: $A^\dagger B=U\in \mathrm{SU(2)}$. We will call such a gate a conditionally controlled gate.
Mathematically, a set of conditionally controlled gates for $U\in\mathrm{SU(2)}$ is defined as
\begin{equation}
    \mathbb{CC}(U) = \{C(A,B)~|~A,B\in\mathrm{SU(2)},~A^\dagger B=U\}.
\end{equation}
Then, using the following gates, a $\mathbb{CC}(U)$ can be synthesized by performing a subgroup-guided search (cf. Sec. \ref{ssec:sds}).
\begin{equation} \label{eq:vxvxd}
    \begin{pmatrix}
        V_x&\\
        &V_x^\dagger
    \end{pmatrix}=
    H_{(1)}
    \begin{pmatrix}
        V_z&\\
        &V_z^\dagger
    \end{pmatrix}
    H_{(1)}.
\end{equation}
\begin{equation} \label{eq:vyvyd}
    \begin{pmatrix}
        V_y&\\
        &V_y^\dagger
    \end{pmatrix}=
    \begin{pmatrix}
    S\\&S^\dagger
\end{pmatrix}H_{(1)}
    \begin{pmatrix}
        V_z&\\
        &V_z^\dagger
    \end{pmatrix}
    H_{(1)}\begin{pmatrix}
    S\\&S^\dagger
\end{pmatrix}^\dagger.
\end{equation}
\begin{equation} \label{eq:vzvzd}
    \begin{pmatrix}
        V_z&\\
        &V_z^\dagger
    \end{pmatrix}=CNOT_{(0,1)}V_{z~(1)}CNOT_{(0,1)}.
\end{equation}
\begin{figure}
    \centering
    \includegraphics[width=10cm]{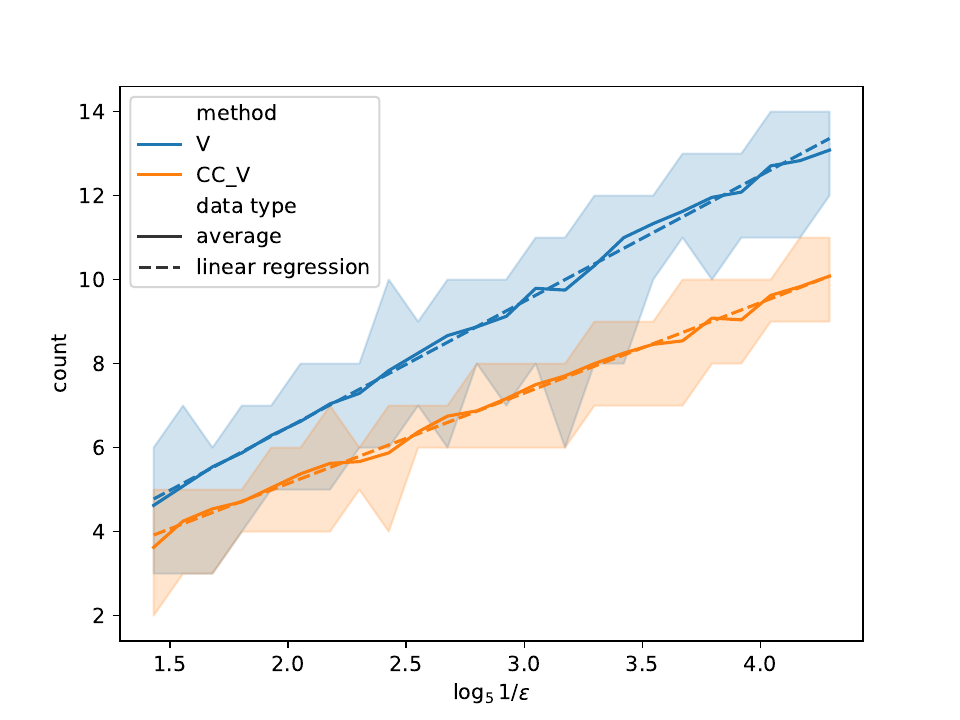}
    \caption{V-count for ten Haar random target SU(2)s for each accuracy $\epsilon$. Blue lines show the results of synthesizing the targets and orange lines show those of synthesizing the CC(target)s. Solid lines are the averages of the individual results, dashed lines are linear regressions of each result, and shaded regions indicate the range of V-count from min to max. The slopes of the blue and orange dashed lines are 3.00 and 2.15.}
    \label{fig:compare}
\end{figure}

\subsection{Meet-in-the-Middle Exhaustive Search} \label{ssec:fds}
We constructed a fast direct search algorithm for SU(2) with meet-in-the-middle and kd-tree 
(Alg. \ref{alg:fds}).
The outer loop of this algorithm is iterated $i\in\{0,1,\cdots,\lfloor k/2\rfloor\}$ times, where $k$ is the V-count for the target $T$.
The inner loops are run $O\left(|\mathbb{B}|^i\right)$ times, where $\mathbb{B}$ is a basis considering now.
Since constructing a kd-tree for $N$ points takes $O(N\log N)$ time and the nearest neighbor query takes $O(\log N)$ time on average, the time complexity of Alg. \ref{alg:fds} is
\begin{equation}
    O\left(\sum_{i=0}^{\lfloor k/2\rfloor} |\mathbb{B}|^{i}|\mathbb{S}|\log\left(|\mathbb{B}|^{i}|\mathbb{S}|\right) + |\mathbb{B}|^{i}\log\left(|\mathbb{B}|^{i}|\mathbb{S}|\right)\right)
    =
    O\left(|\mathbb{B}|^{\lfloor k/2\rfloor}|\mathbb{S}|\log\left(|\mathbb{B}|^{\lfloor k/2\rfloor}|\mathbb{S}|\right)\right).
\end{equation}
In some cases, $L_0$ on line 34 
can be reduced.
The $\mathbb{B}=$ V-basis is a good example.
In this case, $V_i^\dagger V_i=\mathbb{I}$ and $V_iV_i^\dagger=\mathbb{I}$.
Thus, $\mathbb{B}L_0$ on line 34 reduces to
\begin{equation}
    \{BL|B\in\mathbb{B},~L\in L_1,~B(\mbox{the last basis multiplied with }L)\neq \mathbb{I} \}
\end{equation}
when $i>0$,
and hence, the time complexity of Alg. \ref{alg:fds} can be expressed as $O\left((|\mathbb{B}|-1)^{\lfloor k/2\rfloor}|\mathbb{S}|\log\left((|\mathbb{B}|-1)^{\lfloor k/2\rfloor}|\mathbb{S}|\right)\right)$.
Since Alg. \ref{alg:fds} returns an optimal synthesis and the V-count is $3\log_5(1/\epsilon)$ on average, the time complexity on average is
\begin{equation}
    O\left(\frac{\log(1/\epsilon)}{\epsilon^{1.5}}\right).
\end{equation}

Alg. \ref{alg:fds} is generalized to many kinds of targets
such as qudit or multi-qubit gates.
Although kd-tree is easy to implement, it does not scale well to high‑dimensional spaces.
Thus, we should compromise the exactness of the NNS and use ANNS data structures that remain effective in such spaces (e.g. HNSW \cite{malkov2018efficient}, NSG \cite{FuNSG17}).
In SU(d) spaces other than SU(2), we are not able to use Eq. \ref{eq:su2_4d}.
Instead, a simple embedding of SU($d$) in $2d^2$-dimensional Euclidean space can be used because the phase-invariant norm and the norm defined below give almost the same results in $0<\epsilon\ll1$:
\begin{equation}
    \min_{j\in\mathbb{Z}}\sqrt{\frac{\mathrm{tr}\left(\left(e^{2\pi ij/d}U-V\right)\left(e^{2\pi ij/d}U-V\right)^\dagger\right)}{2d}} =
    \min_{j\in\mathbb{Z}}\sqrt{1-\frac{\mathrm{Re}\left(e^{2\pi ij/d}\mathrm{tr}\left(UV^\dagger\right)\right)}{d}}.
\end{equation}

\begin{algorithm}
\caption{Meet-in-the-Middle Exhaustive Search}\label{alg:fds}
\begin{algorithmic}[1]
\Require $0<\epsilon<1 \in \mathbb{R},~T\in\mathrm{SU(2)}, ~ \mathbb{B}, ~ \mathbb{S}$
\Comment{$T$: target unitary, $\mathbb{B}$: basis, $\mathbb{S}$: suffixes}
\Ensure $|T-B_1B_2\cdots B_kS|<\epsilon$
\Comment{$B_1B_2\cdots B_k S$: optimal synthesis}
\State $L_0 \gets \{\mathbb{I}\}$
\State $L_1 \gets \emptyset$
\State $i\gets 0$

\Loop
\State $K\gets$ kd-tree constructed with $L_0\mathbb{S}$
\State
\Comment{kd-tree is constructed in 4D Euclidean space using Eq. \ref{eq:su2_4d}.}
\State $e\gets 1$

\For{$L\in L_1$}
\State $e_1, id_1\gets$ the nearest point to $L^\dagger T$ in $K$
\Comment{$e_1$: diamond distance between $L^\dagger T$ and $L_0\mathbb{S}$}
\State
\Comment{$id_1$: something to identify the nearest point}
\If{$e_1<\epsilon$}
\State $e\gets e_1$
\State Set $B_1,B_2,\cdots,B_{i-1}$ to what correspond to $L$
\State Set $B_{i},B_{i+1},\cdots,B_{2i-1}$ and $S$ to what correspond to $id_1$
\State BREAK
\EndIf
\EndFor
\If{$e<\epsilon$}
\State BREAK
\EndIf

\For{$L\in L_0$}
\State $e_0, id_0\gets$ the nearest point to $L^\dagger T$ in $K$
\If{$e_0<\epsilon$}
\State $e\gets e_0$
\State Set $B_1,B_2,\cdots,B_{i}$ to what correspond to $L$
\State Set $B_{i+1},B_{i+2},\cdots,B_{2i}$ and $S$ to what correspond to $id_0$
\State BREAK
\EndIf
\EndFor
\If{$e<\epsilon$}
\State BREAK
\EndIf

\State $L_1 \gets L_0$
\State $L_0\gets \mathbb{B}L_0$ \label{line:fds-B}
\State $i\gets i+1$
\EndLoop
\end{algorithmic}
\end{algorithm}

\subsection{Subgroup-Guided Search} \label{ssec:sds}
This method is created in order to synthesize conditionally controlled gates.
Let us consider synthesizing one of $\mathbb{CC}(T)$ for some $T\in\mathrm{SU(2)}$.
The synthesized gate $C(A, B)$ should meet the condition $A^\dagger B=T$.
Using the V-basis and Eq. (\ref{eq:vxvxd}-\ref{eq:vzvzd}), the following decomposition is available:
\begin{equation*}
    C(V_1^{i_{1,1}}, V_1^{i_{1,2}})C(V_2^{i_{2,1}}, V_2^{i_{2,2}})\cdots C(V_
    n^{i_{n,1}}, V_n^{i_{n,2}})\in\mathbb{CC}(U),
\end{equation*}
\begin{equation}\label{eq:sgsd}
    (V_1,V_2,\cdots,V_n\in \{V_x, V_y, V_z\},
    ~ i_{1,1}, i_{1,2},\cdots,i_{n,1}, i_{n,2}\in \{\pm 1\}).
\end{equation}
In this case, the restriction can be expressed as
\begin{equation}
    V_n^{-i_{n,1}}\cdots V_1^{-i_{1,1}}V_1^{i_{1,2}}\cdots V_n^{i_{n,2}} = T.
\end{equation}
Then, we can use the meet-in-the-middle technique and kd-tree algorithm together to perform a search (Alg. \ref{alg:sds}) like the one in Sec. \ref{ssec:fds}.
Here, the outer loop iterates over $i\in\{0,1,\cdots,\lfloor k/2\rfloor\}$, while the inner loops run $O(2^i|\mathbb{B}|^i)$ times.
Considering the time complexity for kd-tree, the whole time complexity of Alg. \ref{alg:sds} is
\begin{equation}
    O\left(\sum_{i=0}^{\lfloor k/2\rfloor}2^i|\mathbb{B}|^i\log \left(2^i|\mathbb{B}|^i\right) + 2^i|\mathbb{B}|^i\log\left(|\mathbb{B}|^i\right)\right)
    = O\left({\lfloor k/2\rfloor}(2|\mathbb{B}|)^{\lfloor k/2\rfloor}\log\left(2|\mathbb{B}|\right)\right).
\end{equation}
In cases like the $\mathbb{B}=$ V-basis (c.f. Sec. \ref{ssec:fds}), the time complexity can be reduced to $O\left({\lfloor k/2\rfloor}(2|\mathbb{B}|-1)^{\lfloor k/2\rfloor}\log\left((2|\mathbb{B}|-1)\right)\right)$.

\begin{algorithm}
\caption{Subgroup-guided search}\label{alg:sds}
\begin{algorithmic}[1]
\Require $0<\epsilon<1 \in \mathbb{R},~T\in\mathrm{SU(2)}, ~ \mathbb{B}$
\Comment{$T$: target unitary, $\mathbb{B}$: basis}
\Ensure $|T-B_k^{-i_{n,1}}\cdots B_1^{-i_{1,1}}B_1^{i_{1,2}}\cdots B_k^{i_{n,2}}|<\epsilon$
\State
\Comment{$C(B_1^{i_{1,1}},B_1^{i_{1,2}})\cdots C(B_k^{i_{k,1}},B_k^{i_{k,2}})$: optimal synthesis}
\State $L_0 \gets \{\mathbb{I}\}$
\State $L_1 \gets \{(\mathbb{I}, ~\mathbb{I})\}$
\State $L_2 \gets \emptyset$
\State $i\gets 0$

\Loop
\State $K\gets$ kd-tree constructed with $L_0$
\State
\Comment{kd-tree is constructed in 4D Euclidean space using Eq. \ref{eq:su2_4d}.}
\State $e\gets 1$

\For{$(L_l, L_r)\in L_2$}
\State $e_2, id_2\gets$ the nearest point to $L_l^\dagger TL_r^\dagger$ in $K$
\Comment{$e_2$: distance between $L_l^\dagger TL_r^\dagger$ and $L_0$}
\State
\Comment{$id_2$: something to identify the nearest point}
\If{$e_2<\epsilon$}
\State $e\gets e_2$
\State Set $B_1,B_2,\cdots,B_{i-1}$ to what correspond to $(L_l, L_r)$
\State Set $B_{i},B_{i+1},\cdots,B_{2i-1}$ to what correspond to $id_2$
\State Set $i_{1,1},i_{1,2},\cdots,i_{2i-1,1},i_{2i-1,2}$ to what correspond to $(L_l, L_r)$, and $id_2$
\State BREAK
\EndIf
\EndFor
\If{$e<\epsilon$}
\State BREAK
\EndIf

\For{$(L_l, L_r)\in L_1$}
\State $e_1, id_1\gets$ the nearest point to $L_l^\dagger TL_r^\dagger$ in $K$
\Comment{$e_1$: distance between $L_l^\dagger TL_r^\dagger$ and $L_0$}
\State
\Comment{$id_1$: something to identify the nearest point}
\If{$e_1<\epsilon$}
\State $e\gets e_1$
\State Set $B_1,B_2,\cdots,B_{i-1}$ to what correspond to $(L_l, L_r)$
\State Set $B_{i},B_{i+1},\cdots,B_{2i-1}$ to what correspond to $id_1$
\State Set $i_{1,1},i_{1,2},\cdots,i_{2i-1,1},i_{2i-1,2},j_1,j_2$ to what correspond to $(L_l, L_r)$, and $id_1$
\State BREAK
\EndIf
\EndFor
\If{$e<\epsilon$}
\State BREAK
\EndIf

\State $L_0 \gets \{B L_0 B, B^\dagger L_0B|~B\in\mathbb{B\}}$
\State $L_2 \gets L_1$
\State $L_1\gets \{(L_lB^\dagger,BL_r),(L_lB,BL_r)|~(L_l,L_r)\in L_1,~B\in\mathbb{B}\}$ \label{line:sds-B}
\State $i\gets i+1$
\EndLoop
\end{algorithmic}
\end{algorithm}

\subsection{Controlled gate synthesis}
By combining the results in Sec. \ref{ssec:cgs} and \ref{ssec:sds}, we constructed an algorithm that synthesizes controlled gates.
Since the equation
\begin{equation}
    \begin{pmatrix}
        A\\&B
    \end{pmatrix}
    =\begin{pmatrix}
        BB^{\prime\dagger}\\&BB^{\prime\dagger}
    \end{pmatrix}\begin{pmatrix}
        A^\prime\\&B^\prime
    \end{pmatrix} \quad \left(A,B,A^\prime,B^\prime\in\mathrm{SU(2)}\right)
\end{equation}
holds if and only if $A^\dagger B=A^{\prime\dagger} B^\prime$, the following protocol can synthesize the generalized controlled gates of the form $C(A, B)$.
\begin{enumerate}
    \item Synthesize $C(A^\prime, B^\prime)$ that satisfies $A^\dagger B=A^{\prime\dagger} B^\prime$ by using the subgroup-guided search (Sec. \ref{ssec:sds});
    \item Synthesize $BB^{\prime\dagger}\in\mathrm{SU(2)}$ using any SU(2) synthesizing algorithm.
\end{enumerate}
This protocol can be further generalized to multi-qubit-controlled gate synthesis in Sec. \ref{sec:mcgsa}.The basic idea is that the total degrees of freedom of the $2^{n-1}-1$ subsets $\mathbb{CC}(U)$ together with a single SU(2) add up to $3\cdot2^{n-1}$, which is exactly the same as that of an $n$-qubit generalized controlled gate. Thus, by providing an isomorphism between these two structures, one obtains an efficient decomposition.

Hereafter, we will consider the controlled gate in the narrow sense by assuming that $A$ is the identity.
In this case, from the result of the subgroup-guided search (Eq. (\ref{eq:sgsd})), we can use the following Clifford+V synthesis as the result of the second procedure of the protocol.
\begin{equation} \label{eq:bbdv}
    BB^{\prime\dagger}=V_n^{-i_{n,1}}\cdots V_2^{-i_{2,1}}V_1^{-i_{1,1}}.
\end{equation}
Looking at Fig. \ref{fig:compare}, the blue line, which shows the result of the conditionally controlled gate synthesis using the subgroup-guided search, is below the orange line, which shows the result of the optimal SU(2) gate synthesis.
This implies that Eq. (\ref{eq:bbdv}) sets a severe upper bound $3\log_{\phi}(1/\varepsilon)$ on the V-count of $BB^{\prime\dagger}$. Detail of this upper bound is summerized at Sec. \ref{sec:vcount_sgs}.
We also performed test calculations on this protocol.

\begin{figure}
    \centering
    \includegraphics[width=0.9\linewidth]{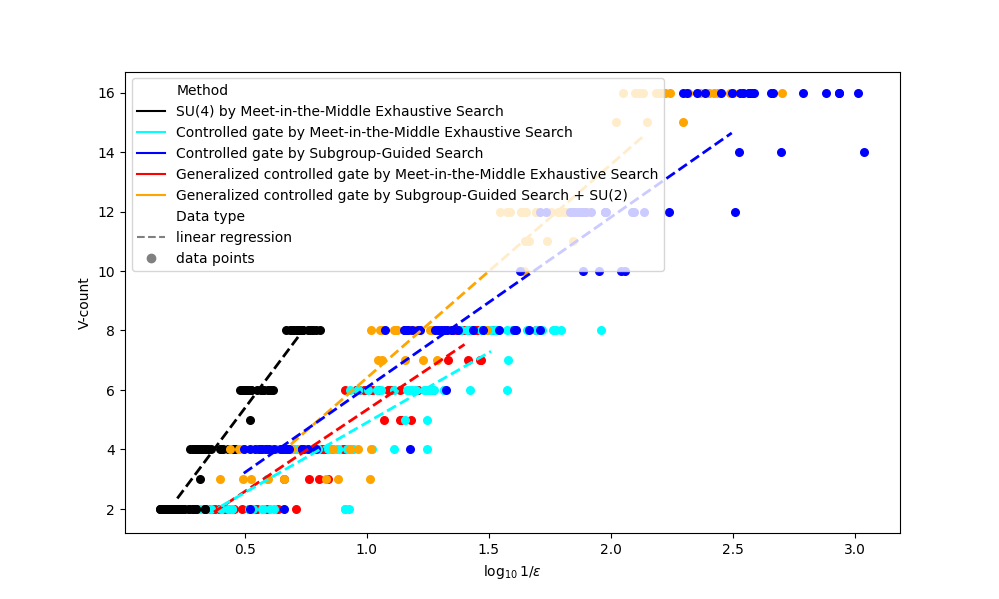}
    \caption{V-count scaling of different synthesis strategies as a function of the target accuracy, plotted against $\log_{10}(1/\varepsilon)$. Dots show individual data points and dashed lines show linear regressions to those averages.
    For 24 Haar-random targets, we fixed the V-count and computed the minimum error $\varepsilon$ for each, repeating this procedure for several even values of the V-count.
    Black - Result of Meet-in-the-Middle Exhaustive Search for targets in SU(4).
    Cyan - Result of Meet-in-the-Middle Exhaustive Search for targets in SU(4) controlled gate.
    Blue - Result of Subgroup-Guided Search for targets in SU(4) generalized controlled gate.
    Red - Result of Meet-in-the-Middle Exhaustive Search for targets in SU(4) generalized controlled gate.
    Orange - Result of Subgroup-Guided Search and SU(2) synthesis search for targets in SU(4) generalized controlled gate.
    Each point with an odd V-count indicates that a sequence of the corresponding odd number of V gates can approximate the target more accurately than a sequence with one additional V gate.
    Each slope and intercept of the linear fits are summarized in Table \ref{tab:comparison}.
    }
    \label{fig:compare_methods}
\end{figure}

\begin{table}[t]
    \centering
    \scriptsize
    \begin{tabular}{|c||c||c|c|c|} \hline
        Synthesis method & Target & Run time & Linear V-count fit & Heuristic V-count estimate \\
        \hline\hline
        Mukhopadhyay\cite{PhysRevA.109.052619} & \multirow{2}{*}{SU($2^N$)} & $O((2\cdot 4^N-3)^{V_\varepsilon})$ & - & - \\
        \cline{1-1} \cline{3-5}
        Meet-in-the-Middle &  & $\tilde{O}((2\cdot 4^N-3)^{V_\epsilon/2})$ & - & - \\
        \hline
        \begin{tabular}{c}Subgroup-Guided\\ + SU(2)\end{tabular} & \begin{tabular}{c}SU($2^N$) generalized\\ controlled gates\end{tabular} & $\tilde{O}(2^{2.5N}\varepsilon^{-1.5})^{*2}$ & - & $3(\log_5(2^{N-2}/\varepsilon)+(2^N-1)\log_{\phi}(2^{N-2}/\varepsilon))+O(1)$ \\
        \hline
        Meet-in-the-Middle & SU(4) & $\tilde{O}(\varepsilon^{-8.02})^{*1}$ & $16.0\log_{29}(1/\varepsilon)-0.10$ & - \\
        \hline
        Meet-in-the-Middle & \multirow{2}{*}{SU(4) controlled gates} & $\tilde{O}(\varepsilon^{-3.44})^{*1}$ & $4.69\log_{\phi}(1/\varepsilon)+0.19$ & - \\
        \cline{1-1} \cline{3-5}
        Subgroup-Guided &  & $\tilde{O}(\varepsilon^{-1.5})^{*2}$ & $5.69\log_{\phi}(1/\varepsilon)+0.36$ & $6\log_{\phi}(1/\varepsilon)+O(1)$ \\
        \hline
        Meet-in-the-Middle & \multirow{2}{*}{\begin{tabular}{c}\raisebox{-1.5ex}{SU(4) generalized} \\ controlled gates\end{tabular}} & $\tilde{O}(\varepsilon^{-4.01})^{*1}$ & $5.46\log_{\phi}(1/\varepsilon)-0.16$ & - \\
        \cline{1-1} \cline{3-5}
        \begin{tabular}{c}Subgroup-Guided\\ + SU(2)\end{tabular} &  & $\tilde{O}(\varepsilon^{-1.5})^{*2}$ & $7.14\log_{\phi}(1/\varepsilon)-0.77$ & \begin{tabular}{c}$3(\log_5(1/\varepsilon)+\log_{\phi}(1/\varepsilon))+O(1)$ \\ $\simeq 7.27\log_{\phi}(\varepsilon)+O(1)$ \end{tabular} \\
        \hline
    \end{tabular}
    \caption{Comparison of the previous study \cite{PhysRevA.109.052619}, Meet-in-the-Middle Exhaustive Searches, and Subgroup-Guided Search. $\phi$ is equal to $5+2\sqrt{6}\simeq9.9$. Each run time is written in $O$ or $\tilde{O}$ notation. The $\tilde{O}$ notation is a variant of the $O$ notation that suppresses polylogarithmic factors and $V_\varepsilon$ is the V-count for SU($2^N$). $V_\varepsilon$ strongly depends on $\varepsilon$, since there exists a lower bound $(4^N-1)\log_{2\cdot4^N-3}(1/\varepsilon)$. Run times with $*1$ are calculated from the linearly fitted V-count, while run times with $*2$ are calculated with a formula derived from the algorithm’s design. Linear V-count fits are results from Figure \ref{fig:compare_methods}. Heuristic V-counts are calculated under the assumption that unitaries generated from the 12 bases used in the subgroup-guided search prevail uniformly. Further details are provided in Sec. \ref{sec:vcount_sgs}.}
    \label{tab:comparison}
\end{table}

\section*{Summary and Discussion}
We constructed two algorithms, a meet-in-the-middle exhaustive search and a subgroup-guided search, based on NNS/ANNS and the meet-in-the-middle technique. Simply combining these two techniques will lead to the meet-in-the-middle exhaustive search and make the run time quadratically shorter than that of a simple exhaustive search (see the upper two run times in Table \ref{tab:comparison}). Considering the subgroup $\mathbb{CC}(U)$ will lead to the subgroup-guided search that further improves the run time for compiling controlled gates (Table \ref{tab:comparison}).
We performed test calculations to show how many gates are actually needed to synthesize target gates (Fig. \ref{fig:compare_methods}).
The slopes and intercepts for these lines are listed as linear V-count fits in Table \ref{tab:comparison}.
For SU(4) generalized controlled targets under the meet-in-the-middle exhaustive search, the fitted slope is approximately $5.46\log_{\phi}(1/\varepsilon)$ (intercept around $-0.16$). In contrast, the subgroup-guided search followed by an SU(2) residual step yields $7.14\log_{\phi}(1/\varepsilon)$ (intercept around $-0.77$) for generalized controlled gates. Although the introduction of $\mathbb{CC}(U)$ degrades the V-count scale, it improves the run time from $\tilde{O}(\varepsilon^{-4.01})$ to $\tilde{O}(\varepsilon^{-1.5})$. Thus, it substantially reduces the time-to-accuracy and pushes the run-time-V-count Pareto front while keeping the synthesis feasible.

The central idea is to reduce generalized controlled-gate synthesis to a “one-shot search over $\mathbb{CC}(U)$ + SU(2) residual cleanup.” The family $\mathbb{CC}(U)$ forms a structured subgroup inside SU(4) (and more generally SU($2^n$)). Restricting the search to this subgroup provides (i) an effective reduction in search dimensionality, (ii) better compatibility with NNS/ANNS data structures, and (iii) improved balance between V-count and runtime. Conceptually, this yields a new design principle for multi-qubit compilation: first conquer an easily reachable subgroup, then seal the gap in a lower-dimensional residual space—distinct from both full-space direct search and naive one-qubit reductions.

The isomorphism between $\mathbb{CC}(U)^{2^{n-1}-1}\times\mathrm{SU}(2)$ and $n$-qubit generalized controlled gates leads to a generalization of the above idea to the multi-qubit case.
This suggests a potential “subgroup-first” trend in multi-qubit compilation. Notably, the $\tilde{O}(\varepsilon^{-1.5})$ design value provides a theoretical foothold for limiting exponent blow-up as we scale from SU(4) to SU($2^n$).

\section*{Acknowledgment}
SY is supported by JSPS KAKENHI Grant No.~JP24KJ0907 and the Forefront Physics and Mathematics Program to Drive Transformation (FoPM).
SA is partially supported by JST PRESTO Grant No.JPMJPR2111, JST Moonshot R\&D MILLENNIA Program (Grant No.JPMJMS2061), JPMXS0120319794, and CREST (Japan Science and Technology Agency) Grant No.JPMJCR2113.

\appendix
\section{Multi-qubit-controlled gate synthesis algorithm} \label{sec:mcgsa}
Before we treat $n>2$-qubit gates, we define $C_{n-1}(A, B, i)$ and $\mathbb{CC}_{n-1}(U, i)~(A,B,U\in\mathrm{SU(2)},~i\in\mathbb{Z},0\leq i<2^{n-1})$ as
\begin{equation}
    C_{n-1}(A_0, A_1, i) = C(A_{i\star 0},\cdots,A_{i\star j},\cdots,A_{i\star (2^{n-1}-1)}),
\end{equation}
\begin{equation}
    \mathbb{CC}_{n-1}(U, i)=\{C_{n-1}(A,B,i)~|~A,B\in\mathrm{SU(2)},~A^\dagger B=U\}
\end{equation}
where $(\star)$ operator is the inner product over a vector space on $\mathbb{F}_2$. (e.g. $23\star 30 = (10111)_2\star(11110)_2 = \mathrm{parity}((10110)_2) = 1$.)
$C_{n-1}(V_j,V_j^\pm,i)$ can be constructed by conjugating $C(V_j,V_j^\pm)$ with CNOT gates.
Below, we show how the $n$-qubit controlled gate $C(A_0,A_1,\cdots,A_{2^n-1})$ is decomposed into the product $\prod_{i=0}^{2^n-1}C_n(U_{i,0}, U_{i,1}, i)$ by considering the following functions for $i=1,\cdots,2^n-1$.
\begin{eqnarray}
    \alpha_{n;i} : \quad & \mathrm{SU(2)}^{2\cdot 2^n-1-i} \to \mathrm{SU(2)},\\
    & (A_0,A_1,\cdots,A_{2^n-1},U_{2^n-1,0},\cdots,U_{i+1,0}) \mapsto U_{i,0}^\dagger U_{i,1}.
\end{eqnarray}
This definition means that
\begin{equation}
    C_n(U_{i,0},U_{i,1},i)\in\mathbb{CC}_n(\alpha_{n;i},i)
\end{equation}
and
\begin{equation}
    U_{0,0}=A_0\prod_{i=1}^{2^n-1}U_{2^n-i,0}^\dagger.
\end{equation}
Also, the codomains of $\alpha_{n;i}$ imply that the whole algorithm to compile $C(A_0,A_1,\cdots,A_{2^n-1})$ decides $U_{i,0}$ and $U_{i,1}$ in order from $i=2^n-1$ to $i=1$.
Therefore, the algorithm proceeds by iterating the following processes over $i = 2^n - 1, \cdots, 1$ and finally synthesizing $U_{0;0}$:
\begin{enumerate}
    \item Calculate $U_{i,0}$ and $U_{i,1}$;
    \item Synthesize one element of $\mathbb{CC}_n(\alpha_{n;i}, i)$ by using Alg. \ref{alg:sds}.
\end{enumerate}

In what follows, we determine $\alpha_{n;i}$ inductively by assuming that $\alpha_{n-1;i}$ has already been obtained.
Since
\begin{flalign*}
    &\prod_{i=0}^{2^n-1}C_n(U_{i,0}, U_{i,1}, i) ~=~
    C_n\left(\prod_{i=0}^{2^n-1}U_{i,0},~\_\,,0\right)& 
\end{flalign*}
\begin{equation}
    \left(\prod_{i=1}^{2^n-1} C_n\left(\mathbb{I},~\left(U_{i+1,0}\cdots U_{2^n-1,0}\right)^\dagger U_{i,0}^\dagger U_{i,1}\left(U_{i+1,0}\cdots U_{2^n-1,0}\right), \,i\right)\right),
\end{equation}
there are equivalent functions $\beta_{n;i}$ to $\alpha_{n;i}$ such that
\begin{eqnarray}
    \beta_{n;i} : \quad & \mathrm{SU(2)}^{2^n} \to \mathrm{SU(2)},\\
    & (A_0,A_1,\cdots,A_{2^n-1}) \mapsto \left(U_{i+1,0}\cdots U_{2^n-1,0}\right)^\dagger U_{i,0}^\dagger U_{i,1}\left(U_{i+1,0}\cdots U_{2^n-1,0}\right)
\end{eqnarray}
and
\begin{equation} \label{eq:aubu}
    \alpha_{n;i}=\left(U_{i+1,0}\cdots U_{2^n-1,0}\right)\beta_{n;i}\left(U_{i+1,0}\cdots U_{2^n-1,0}\right)^\dagger.
\end{equation}
First, we consider $\beta_{n;i}$ for $1\leq i< 2^{n-1}$. 
If we assume that $\beta_{n;i}$ have already been decided for $2^{n-1}\leq i< 2^n$, we can decide $\beta_{n;i}$ for $1\leq i< 2^{n-1}$ with the following:
\begin{equation}
    \beta_{n;i}=
    \beta_{n-1;i}\left(A_0, A_1\left(\prod_{j\in S_{n;1}(1)}\beta_{n;j}\right)^\dagger, \cdots, A_{2^{n-1}-1}\left(\prod_{j\in S_{n;2^{n-1}-1}(1)}\beta_{n;j}\right)^\dagger\right),
\end{equation}
where $S_{n;j}(x)=\{y\in\mathbb{Z}\,|~ 2^{n-1}\leq y<2^n,\,y\star j=x\}$ for $j\in\{0,1,\cdots,2^n-1\},~x\in\{0,1\}$.

Next, let us consider $\beta_{n;i}$ for $2^{n-1}\leq i< 2^n$.
In this case,
\begin{equation} \label{eq:bb}
    \beta_{n;j}\beta_{n;j+1}=\beta_{n-1;j/2}(A_0,A_2,\cdots,A_{2^n-2})~(j\in\{2^{n-1},2^{n-1}+2,\cdots,2^n-2\})
\end{equation}
and we define
\begin{equation} \label{eq:gbb}
    \gamma_{n;k}\left(A_0^\dagger A_{2^{n-1}},~A_1^\dagger A_{2^{n-1}+1},\,\cdots,\, A_{2^{n-1}-1}^\dagger A_{2^n-1}\right)
    = \beta_{n;\,2k+2^{n-1}}^\dagger\beta_{n;\,2k+2^{n-1}+1} ~(k=0,1,\cdots,2^{n-2}-1).
\end{equation}
The $\gamma_{n;k}$ are well defined because $\beta_{n;i}$ for $i\geq2^{n-1}$ depend $A_0, A_1,\cdots,A_{2^n-1}$ only through $A_0^\dagger A_{2^{n-1}},$ $A_1^\dagger A_{2^{n-1}+1},\,\cdots,\,$ $ A_{2^{n-1}-1}^\dagger A_{2^n-1}$.
$\gamma_{n;k}\left(B_0,B_1,\cdots,B_{2^{n-1}-1}\right)$ can be decided as
\begin{equation}
    \gamma_{n;k}=\begin{cases}
        \gamma_{n-1;k-2^{n-3}}\left(\sqrt{B_0B_{2^{n-2}}}B_{2^{n-2}}^\dagger,~\cdots,\,\sqrt{B_{2^{n-2}-1}B_{2^{n-1}-1}}B_{2^{n-1}-1}^\dagger\right)\quad\\
        \qquad\qquad(2^{n-3}\leq k<2^{n-2}) \\
        \gamma_{n-1;k}\left(\left(\displaystyle\prod_{l\in S_{n-2;0}(1)}\beta_{n;l}\right)B_0\left(\displaystyle\prod_{l\in S_{n-2;0}(0)}\beta_{n;l}\right)^\dagger,~\cdots, \right.\\
        \left.\qquad\left(\displaystyle\prod_{l\in S_{n-2;\,2^{n-2}-1}(1)}\beta_{n;l}\right)B_{2^{n-2}-1}\left(\displaystyle\prod_{l\in S_{n-2;\,2^{n-2}-1}(0)}\beta_{n;l}\right)^\dagger\right) \quad \\
        \qquad\qquad(0\leq k<2^{n-3}).
    \end{cases}
\end{equation}
(The $0\leq k<2^{n-3}$ part is will be decided following to the decision of $\beta_{n;i}$ for $i=3\cdot2^{n-2},\cdots,2^n$.)
After deciding $\gamma_{n;k}$, we can decide $\beta_{n;2k+2^{n-1}}$ and $\beta_{n;2k+2^{n-1}+1}$ by using Eq. (\ref{eq:bb}),(\ref{eq:gbb}).
This and Eq. (\ref{eq:aubu}) lead to the decision of all $\alpha_{n;i}$.

To complete the induction, we should mention the base cases $\beta_{1;1}$ and $\gamma_{2;0}$.
They are
\begin{equation}
    \beta_{1;1}\left(A, B\right)=A^\dagger B
\end{equation}
and
\begin{equation}
    \gamma_{2;0}(A,B)=B.
\end{equation}

The above method leads to not only multi-qubit-controlled gate synthesis, but also slightly more generalized gate synthesis considering forms like those of CNOTs$\times\mathbb{CC}\times$CNOTs (i.e. match gates).

\section{V-count of Subgroup-Guided Search} \label{sec:vcount_sgs}
In Subgroup-Guided Search, the following 12 bases appear.
\begin{equation}
    \left\{V_P, V_P^\dagger\middle|P=\mathbb{I}\otimes X, \mathbb{I}\otimes Y, \mathbb{I}\otimes Z, Z\otimes X, Z\otimes Y, Z\otimes Z\right\}, \quad 
    V_P = \frac{\mathbb{I}+2iP}{\sqrt{5}}.
\end{equation}
We consider the sequence of $n$ bases to approximate some target in an allowed error $\varepsilon$. By counting how many distinct products are computed from such sequences, we can set the lower bound on the average V-count. The most naive one is $3\log_{12}1/\varepsilon+O(1)$, because the target resides in a 3 dimensional space and the number of distinct products $N_p\leq 12^n$.
But it can be easily improved to $3\log_{11}1/\varepsilon+O(1)$, because $V_PV_P^\dagger=I$ and $N_P\leq12\cdot 11^{n-1}$.
Considering that there exist commuting pairs of $V_P$ other than $(V_P, V_P)$ and $(V_P, V_P^\dagger)$ (i.e. $(V_{\mathbb{I}\otimes X},V_{Z\otimes X}^\dagger)$, this evaluation can be further improved.

In order to avoid counting multiple sequences that result in the same product due to the commutation relations but differ in the order of $V_P$, we define that a $V_P$ whose first Pauli matrix in $P$ is $I$ precedes that with $Z$ in lexicographic order, and only the lexicographically earlier one is adopted.
To calculate the number of such sequences adopted, we consider two states of them.
\begin{enumerate}
    \item a sequence ends with $V_P$ or $V_P^\dagger$ whose first Pauli matrix in $P$ is $I$
    \item a sequence ends with $V_P$ or $V_P^\dagger$ whose first Pauli matrix in $P$ is $Z$
\end{enumerate}
The number of adopted sequences consisting of $n$ gates can be obtained as the sum of the components of the following vector.
\begin{equation}\label{eq:state_num}
    \begin{pmatrix}
        5 & 4\\
        6 & 5
    \end{pmatrix}^{n-1}\begin{pmatrix}
        6\\6
    \end{pmatrix}.
\end{equation}
The first and second components correspond to the numbers of sequences adopted in the first and second states, respectively.
Thus,
\begin{equation}
    N_P\leq \frac{\sqrt{3}}{\sqrt{2}}((5+2\sqrt{6})^n - (5-2\sqrt{6})^n) \leq \frac{\sqrt{3}}{\sqrt{2}}(5+2\sqrt{6})^n
\end{equation}
and the lower bound is improved to $3\log_\phi 1/\varepsilon + O(1)$ where $\phi=5+2\sqrt{6}$.

\bibliographystyle{alpha}
\bibliography{references}

\begin{thebibliography}{FXWC19}

\bibitem[AKT24]{SA}
Seiseki Akibue, Go~Kato, and Seiichiro Tani.
\newblock {Probabilistic Unitary Synthesis with Optimal Accuracy}.
\newblock {\em ACM Transactions on Quantum Computing}, 5(3), August 2024.

\bibitem[Ben75]{10.1145/361002.361007}
Jon~Louis Bentley.
\newblock Multidimensional binary search trees used for associative searching.
\newblock {\em Commun. ACM}, 18(9):509^^e2^^80^^93517, September 1975.

\bibitem[BGS13]{PhysRevA.88.012313}
Alex Bocharov, Yuri Gurevich, and Krysta~M. Svore.
\newblock {Efficient decomposition of single-qubit gates into $V$ basis circuits}.
\newblock {\em Phys. Rev. A}, 88:012313, Jul 2013.

\bibitem[BK05]{BK05}
Sergey Bravyi and Alexei Kitaev.
\newblock {Universal quantum computation with ideal Clifford gates and noisy ancillas}.
\newblock {\em Phys. Rev. A}, 71:022316, Feb 2005.

\bibitem[Cop02]{C02}
D.~Coppersmith.
\newblock {An approximate Fourier transform useful in quantum factoring}, 2002.

\bibitem[CW12]{AN12}
Andrew~M. Childs and Nathan Wiebe.
\newblock {Hamiltonian simulation using linear combinations of unitary operations}.
\newblock {\em Quantum Information and Computation}, 12(11 \& 12), November 2012.

\bibitem[DCS13]{DS13}
Guillaume Duclos-Cianci and Krysta~M. Svore.
\newblock {Distillation of nonstabilizer states for universal quantum computation}.
\newblock {\em Phys. Rev. A}, 88:042325, Oct 2013.

\bibitem[EK09]{EK09}
Bryan Eastin and Emanuel Knill.
\newblock {Restrictions on Transversal Encoded Quantum Gate Sets}.
\newblock {\em Phys. Rev. Lett.}, 102:110502, Mar 2009.

\bibitem[FXWC19]{FuNSG17}
Cong Fu, Chao Xiang, Changxu Wang, and Deng Cai.
\newblock {Fast Approximate Nearest Neighbor Search With The Navigating Spreading-out Graphs}.
\newblock {\em {PVLDB}}, 12(5):461 -- 474, 2019.

\bibitem[Got98]{G98}
Daniel Gottesman.
\newblock {The Heisenberg Representation of Quantum Computers}, 1998.

\bibitem[GSLW19]{GSLW19}
Andr\'{a}s Gily\'{e}n, Yuan Su, Guang~Hao Low, and Nathan Wiebe.
\newblock Quantum singular value transformation and beyond: exponential improvements for quantum matrix arithmetics.
\newblock In {\em Proceedings of the 51st Annual ACM SIGACT Symposium on Theory of Computing}, STOC 2019, page 193^^e2^^80^^93204, New York, NY, USA, 2019. Association for Computing Machinery.

\bibitem[LC19]{LC19}
Guang~Hao Low and Isaac~L. Chuang.
\newblock Hamiltonian {S}imulation by {Q}ubitization.
\newblock {\em {Quantum}}, 3:163, July 2019.

\bibitem[Muk24]{PhysRevA.109.052619}
Priyanka Mukhopadhyay.
\newblock {Synthesis of V-count-optimal quantum circuits for multiqubit unitaries}.
\newblock {\em Phys. Rev. A}, 109:052619, May 2024.

\bibitem[MY18]{malkov2018efficient}
Yu~A Malkov and Dmitry~A Yashunin.
\newblock Efficient and robust approximate nearest neighbor search using hierarchical navigable small world graphs.
\newblock {\em IEEE transactions on pattern analysis and machine intelligence}, 42(4):824--836, 2018.

\bibitem[NC10]{NielsenChuang}
Michael~A. Nielsen and Isaac~L. Chuang.
\newblock {\em {Quantum Computation and Quantum Information: 10th Anniversary Edition}}.
\newblock Cambridge University Press, 2010.

\end{thebibliography}

\end{document}